\documentclass[12pt]{article}
\usepackage{amsmath}
\usepackage{graphicx,psfrag,epsf}
\usepackage{enumerate}
\usepackage{natbib}
\usepackage{url} % not crucial - just used below for the URL 
\usepackage[figuresright]{rotating}
\usepackage{booktabs}
\usepackage{adjustbox}
\usepackage{setspace}
\usepackage{xcolor}
\usepackage{soul}
\usepackage{caption}
\usepackage{multirow}
\usepackage{lscape}
\usepackage{algorithmic, algorithm}
\newcommand{\blind}{1}

\begin{document}

\def\spacingset#1{\renewcommand{\baselinestretch}%
{#1}\small\normalsize} \spacingset{1}

%%%%%%%%%%%%%%%%%%%%%%%%%%%%%%%%%%%%%%%%%%%%%%%%%%%%%%%%%%%%%%%%%%%%%%%%%%%%%%

\if1\blind
{
  \title{\bf Prioritizing Variables for Observational Study Design using the Joint Variable Importance Plot}
  \author{Lauren D. Liao$^{1,}$\thanks{
    The authors gratefully acknowledge support from \textit{Hellman Fellowship, National Science Foundation 2142146 and DGE 2146752, National Institute of Diabetes and Digestive and Kidney Diseases K01DK120807, National Heart, Lung, and Blood Institute R01HL157666, and Kaiser Permanente Northern California Community Benefits Program RNG209492}. The authors thank {David Bruns-Smith,} Avi Feller, {Erin Hartman,} Melody Y. Huang, {Yaxuan Huang, Sizhu Lu, Arisa Sadeghpour, Andy Shen, and Arnout van Delden for valuable comments.} Lauren D. Liao is a PhD candidate at Division of Biostatistics, Berkeley, CA 94720 (Corresponding Email: \textit{ldliao@berkeley.edu}). Yeyi Zhu is a Research Scientist II (equivalent to associate professor) at Kaiser Permanente Northern California Division of Research, Oakland, CA 94612 and Associate Adjunct Professor at Department of Epidemiology and Biostatistics, University of California, San Francisco, San Francisco, CA 94143 (Email: \textit{yeyi.zhu@kp.org}). Amanda L. Ngo is data reporting and analytics consultant III at Kaiser Permanente Northern California Division of Research, Oakland, CA 94612 (Email: \textit{Amanda.L.Ngo@kp.org}). Rana F. Chehab is a research post doctoral fellow at Kaiser Permanente Northern California Division of Research, Oakland, CA 94612 (Email: \textit{Rana.Chehab@kp.org}). Samuel D. Pimentel is an Assistant Professor at the Department of Statistics, University of California, Berkeley, Berkeley, CA 94720 (Email: \textit{spi@berkeley.edu}).} , Yeyi Zhu$^2$, Amanda L. Ngo$^2$,\\ Rana F. Chehab$^2$, Samuel D. Pimentel$^3$\\
    \\
    $^1$Division of Biostatistics, Berkeley, CA 94720 \\
    $^2$Kaiser Permanente Northern California Division of Research,\\ Oakland, CA 94612 \\
    $^3$Department of Statistics, Berkeley, CA 94720 \\
    $^*$For correspondence email: \textit{ldliao@berkeley.edu} 
    }
    
  \maketitle
} \fi

\if0\blind
{
  \bigskip
  \bigskip
  \bigskip
  \begin{center}
    {\LARGE\bf Prioritizing Variables for Observational Study Design using the Joint Variable Importance Plot}
\end{center}
  \medskip
} \fi

\bigskip
\begin{abstract}
Observational studies of treatment effects {require adjustment for confounding variables.} {However, causal inference methods typically} cannot {deliver perfect adjustment} on all measured baseline variables, and there is often ambiguity about which variables should be prioritized. Standard prioritization methods based on treatment imbalance alone neglect variables' relationships with the outcome. We propose the joint variable importance plot to guide variable prioritization for observational studies. Since not all variables are equally relevant to the outcome, the plot adds outcome associations to quantify the potential confounding jointly with the standardized mean difference. To enhance comparisons on the plot between variables with different confounding relationships, we also derive and plot bias curves. Variable prioritization using the plot can produce recommended values for tuning parameters in many existing matching and weighting
methods. We showcase the use of {the} joint variable importance plots {in the design of  a balance-constrained matched study} to evaluate whether taking an antidiabetic medication, glyburide, increases the incidence of C-section delivery among pregnant individuals with gestational diabetes.
\end{abstract}

\noindent%
\vfill

\newpage
\spacingset{1.45} % DON'T change the spacing!
\section{INTRODUCTION}
\label{sec:intro}

Researchers often seek to evaluate treatments to understand whether they are beneficial. In observational (non-randomized) studies,  {treatments may be confounded, or associated with other baseline variables so that it is unclear whether to attribute group outcome differences to treatment or baseline dissimilarity.  To reliably estimate an effect, researchers must adjust for these variables, typically either by modeling their impact on study outcomes or by creating new comparison groups that eliminate baseline differences or imbalances, for example by matching or weighting.} 

One crucial decision is deciding which variables are most important for adjustment. While {creating comparison groups with} perfect balance on the joint distribution of all baseline variables, {or conditioning appropriately on this joint distribution in an outcome model,} is sufficient to remove observed sources of confounding, this is impossible in datasets with a large number of measured variables. Attempting to adjust for too many variables can lead to undesirable designs, such as {heavily overfitted models,} matches with too few subjects to be useful \citep{zubizarreta2014matching}, or weighting designs with high-variance weights that hurt precision \citep{miratrix2018worth}. 
Many modern causal inference methods are designed with variable prioritization in mind {and} incorporate substantive or data-driven knowledge about which variables are likely to matter most. 
 {These include regularization procedures for outcome regression \citep{athey2018approximate}, balance tolerances for weighting \citep{ben2021varying}, and covariate distances or balancing constraints for matching \citep{stuart2010matching, pimentel2015large, bennett2020building}.} 
 However, there is a need for better data-driven diagnostic tools 
 to guide researcher choices about prioritization.

{Researchers may be tempted to prioritize variables based on standard balance diagnostics, including tables of}
standardized mean differences (SMD) for each variable or Love plots \citep{ahmed2006heart, stuart2011matchit, greifer2020assessing, hansen2008covariate, rosenbaum1985bias}. These diagnostics {are useful for} highlighting variables with large imbalances between treated and control groups. {However, }
prioritizing variables according to their imbalance ignores important information about the role of each variable in the outcome model. Variables strongly related to treatment but unrelated to outcomes are \textit{not} confounders. In contrast, if variables are strongly associated with the outcome but with only moderate imbalance, they may be \textit{strong} confounders. When choosing which baseline variables to prioritize for adjustment, focusing solely on the treatment imbalance can risk ignoring variables that should take precedence due to their outcome importance.

{The joint importance of covariate-treatment and covariate-outcome relationships is a general  principle in observational causal inference, not specific to a particular framework or set of identification assumptions.  For example, outcome regression approaches typically do not make assumptions about the treatment-covariate relationship, but these relationships influence treatment effect estimation (see Section \ref{subsec:bias_curves}).
Similarly, matching and weighting approaches are typically motivated by models of the treatment variable in covariates, 
but similarity of outcomes within matched pairs or across weighted groups affects residual bias \citep{sales2018rebar, ben2021balancing}.  Another reason outcome-covariate relationships matter is their influence on sensitivity to unmeasured bias.  In both matching and weighting, increasing homogeneity of the outcomes via better control for prognostic covariates increases robustness to worst-case confounding as measured by design sensitivity \citep{rosenbaum2005heterogeneity, huang2023design}.  Unfortunately design sensitivity is understudied in observational study design, and diagnostic tools to improve it are badly needed.}

{To meet these needs} we propose selecting high-priority variables for adjustment 
using the joint treatment-outcome variable importance plot (jointVIP). JointVIP represents each variable in two dimensions: one describing treatment model importance as measured by the SMD, and one describing outcome-model importance, measured by outcome correlation among controls from 
{a} pilot sample (chosen disjointly from the analysis sample to ensure the integrity of the analysis). In addition, under a set of simple working models, the bias incurred by ignoring each variable can be derived separately and represented on the plot using unadjusted bias curves, enhancing opportunities for variable comparisons. We show an example comparison between the traditional Love plot and jointVIP with a subset of the baseline variables from the case study (absolute measures shown in Figure \ref{fig:love_compare} and signed measures shown in Supplemental Appendix A.1). 

{We illustrate jointVIP in detail in a case study of drug safety for diabetes medication in pregnant individuals.  Specifically, we use a matched design with refined covariate balance constraints, which require a prioritized list of variables to be specified for balancing, and jointVIP provides a principled way to choose this. However, we argue that jointVIP's value is not specific to a given estimation strategy or set of identification assumptions.}

\begin{figure}[ht]
    \begin{center}
        \includegraphics[scale=0.65]{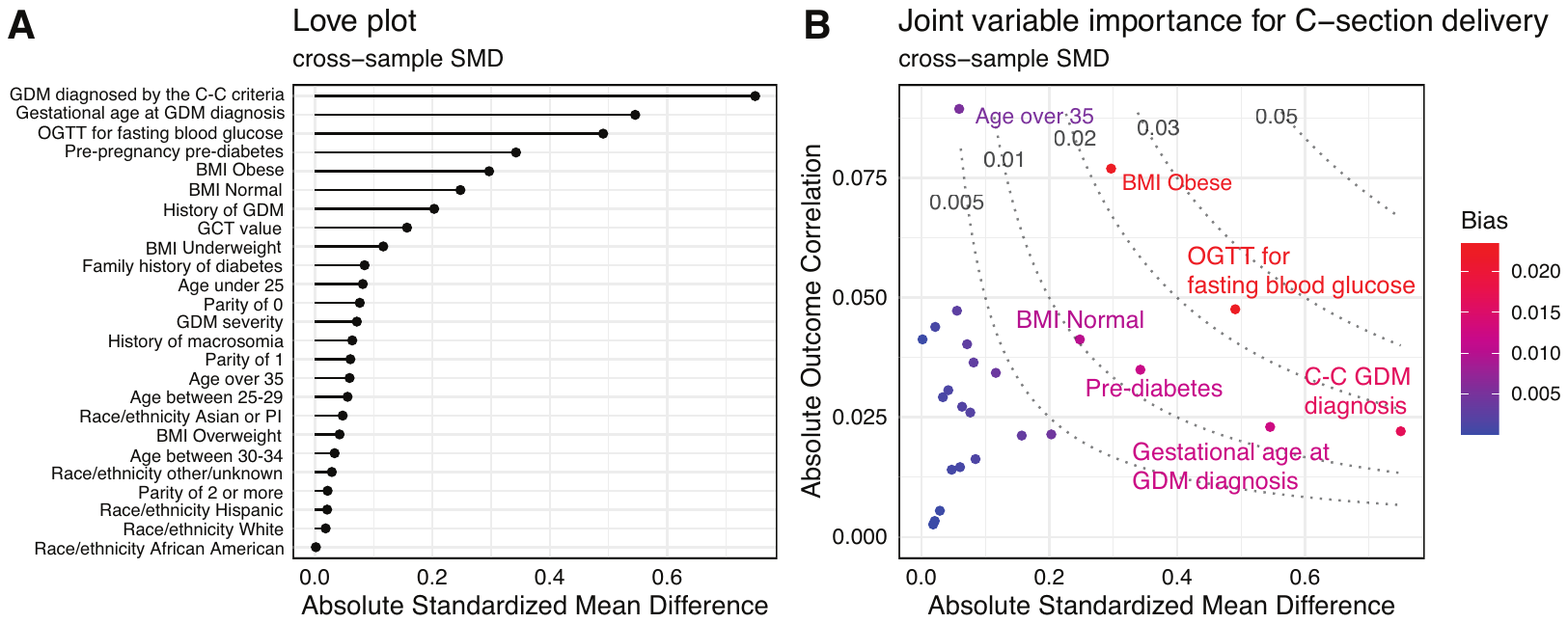}
    \end{center}
    \caption{Comparison between the Love plot and the joint variable importance plot (jointVIP). Note that some variables (BMI in the obese category and OGTT for fasting blood glucose used at GDM diagnosis) take on much more prominent positions in jointVIP than in the Love plot, which only displays SMD values.\\
    \footnotesize {
    BMI: body mass index, C-C: Carpenter-Coustan, C-section: Cesarean section, GDM: gestational diabetes, OGTT: oral glucose tolerance test, PI: Pacific Islander, SD: standard deviation, SMD: standardized mean difference.}}
    \label{fig:love_compare}
\end{figure}
\section{METHOD}
\subsection{Joint variable importance plot construction}
\label{subsec:jointvip_construction}

The high-level purpose of the jointVIP is to illustrate two different dimensions of a variable's possible role as a confounder -- its imbalance, or association with the treatment variable, and its association with the outcome -- on two axes, with each variable plotted as a single point. We now discuss the specific measures of variable importance on each axis.

For treatment model importance, described by the $x$-axis, we use SMDs, or differences between the treated mean and the control mean divided by an estimate of the variable's standard deviation. Many different standard deviation estimates have been proposed leading to slightly different SMD definitions; we focus on a version denoted as the ``cross-sample" SMD, which uses the sample standard deviation of the variable in question computed in the pilot (control) sample. For more motivation and discussion of the cross-sample SMD, see Section \ref{subsec:bias_curves}. The variant we propose is similar to an effect size estimator from \citet{glass1976primary}, which standardizes the mean difference by dividing by the standard deviation from the control group \citep{hedges1981distribution}. SMDs allow intuitive comparisons across variables with very different scales, including both binary and continuous variables. 
They are widely used to assess imbalance and are commonly reported in balance tables or Love plots. Thus, using SMD on the $x$-axis allows jointVIP to preserve all information typically contained in the Love plot while adding new insights.

For outcome model importance, represented on the $y$-axis, we compute the sample Pearson correlation between each variable and the outcome among 
controls.
Sample correlation is a familiar, bounded quantity and makes sense  for relationships not only between two continuous variables but also between two binary variables (phi coefficient), and between binary and continuous variables (point biserial correlation) \citep{pearson1895vii}. The outcome relationship is calculated only among controls to avoid having to model treatment effects. 

{It is vital that the outcome correlations be computed in a pilot sample separate from the data used for the ultimate outcome analysis.} Using controls from the analysis sample for computing outcome correlations can bias treatment effect estimates. For example, suppose treated and control samples exhibit imbalance on several continuous background variables (with treated individuals taking larger values), but the study outcome is independent of all these variables in the population. If we compute sample outcome correlations in the analysis control sample and {form matched pairs based solely on} the variable with the largest such (positive) sample correlation, we  essentially match on the variable with the largest spurious correlation (with the random outcome noise in the current sample). Because of the imbalance, the matching algorithm will systematically select controls with large values for the spuriously correlated variable. Hence, the result will have large positive outcome errors that introduce positive bias into the average outcome for the matched controls. For {related examples} see 
\citet{hansen2008prognostic} and 
\citet{abadie2018endogenous}.

{To construct a pilot sample, one may select}
a small (10-20\%) portion of the control sample at random from the full control group. To ensure the pilot sample is drawn from the portion of the control space most relevant for the observational study, \citet{aikens2020pilot} instead suggest conducting an initial round of matching on a standard Mahalanobis distance to pair each treated subject to two controls, then selecting one control from each set at random to construct the pilot sample. Alternatively, external data separate from the analysis of interest may be used as a pilot sample.

\subsection{Addition of unadjusted bias curves for variable comparison}
\label{subsec:bias_curves}

Comparing {the relative importance of} two distant points on the jointVIP, one with a high outcome correlation and low SMD, and the other with a high SMD and low outcome correlation,  can be difficult.   
A natural answer lies in the relative sizes of the biases contributed by ignoring each variable, since our ultimate goal is to avoid biases in treatment effect estimation. We consider each baseline variable and evaluate the bias incurred by omitting this potential confounder under {a simple one-variable model}. Inspired by \citet{cinelli2020making} and \citet{soriano2021interpretable}, we plot these bias estimates {as curves}
on the jointVIP.

For any baseline variable $X_j$ with  $j \in {1,...,J}$), consider the sample least-squares fit of outcome $Y$ on baseline variable $X_j$ and binary treatment $Z$:
\begin{align}
Y&= Z\tau_0+X_j\beta_j+\hat{\epsilon} \label{eq:eq1}
\end{align}
Here $\hat{\epsilon}$ is a residual.
In addition, consider two related sample regressions:
\begin{align}
Y &= Z \tau + \hat{e}\label{eq:eq2}\\
X_j &= Z\Delta_j+\hat{u}\label{eq:eq3}
\end{align}
Following Cochran's formula \citep{cox2007generalization}, we may use \eqref{eq:eq3} to
rewrite \eqref{eq:eq1} and obtain a new representation for \eqref{eq:eq2}:
\begin{align}
\begin{split}\label{eq:eq4}
Y {}=X_j\beta_j+Z\tau_0+\hat{\epsilon}
&= (Z\Delta_j+\hat{u})\beta_j+Z\tau_0+\hat{\epsilon}%\\
%&= Z\Delta_j\beta_j+\hat{u}\beta_j+Z\tau_0+\hat{\epsilon}
= Z(\Delta_j\beta_j+\tau_0)+(\hat{u}\beta_j+\hat{\epsilon})
\end{split}
\end{align}
Since the new error term $(\hat{u}\beta_j + \hat{\epsilon})$ is orthogonal to $Z$ by the construction of residuals $\hat{u}$ and $\hat{\epsilon}$,
we have $\tau = (\Delta_j\beta_j+\tau_0)$ and $\hat{e} = \hat{u}\beta_j+\hat{\epsilon}$. Note that until now we have made no model assumptions, merely fit regressions using sample quantities; however, if we add a working assumption that {triples $(X,Y,Z)$ are sampled independently from an infinite population, with} model \eqref{eq:eq1} %\st{is} 
correctly specified (i.e. that $E(Y|X_j,Z) = \beta^{pop}_jX_j + \tau^{pop}Z$ for some parameters $\beta_j^{pop}$ and $\tau^{pop}$), then the difference
\begin{align}
\tau - \tau_0 &= \Delta_j\beta_j\label{eq:eq6}
\end{align} is an estimate of the large-sample omitted variable bias (OVB) incurred by estimating treatment effects via regression on $Z$ alone, ignoring $X_j$.

Importantly for our purposes, the OVB can be rewritten in terms of sample correlation between $X_j$ and $Y$ and a SMD with normalization by the control standard deviation. The key is that when equation \eqref{eq:eq1} is fit on controls alone (as it will be in our pilot-sample approach), both \eqref{eq:eq1} and \eqref{eq:eq3} are simple regressions. We  rewrite the corresponding simple regression equations using familiar simple regression formulae. $S_{Y_{pilot}}$ and $S_{X_{j,pilot}}$ denote the standard deviation of the pilot sample for outcome and the standard deviation of the confounder in question respectively. We include the \textit{pilot} and \textit{analysis} notations for clarity.
\begin{align}
\beta_{j}&=r_{X_{j,pilot},Y_{pilot}}\frac{S_{Y_{pilot}}}{S_{X_{j,pilot}}}\label{eq:eq8}\\
\Delta_j &= \Bar{X}_{j1, analysis}-\Bar{X}_{j0, analysis}\label{eq:eq11}
\end{align}
Using \eqref{eq:eq3}, we obtain \eqref{eq:eq11}, where $\Bar{X}_{j1, analysis}$ and $\Bar{X}_{j0, analysis}$ denote variable $j$'s sample means among treated subjects and controls, respectively, in the analysis sample. Substituting into expression \eqref{eq:eq6} and rearranging, we obtain:
\begin{align}
\frac{\Delta_j\beta_j}{S_{Y_{pilot}}} &= r_{X_{j,pilot},Y_{pilot}}\frac{(\Bar{X}_{j1,analysis}-\Bar{X}_{j0,analysis})}{S_{X_{j,pilot}}}\label{eqn:unadj_bias}
\end{align}
The left-hand side is a conveniently normalized version of the OVB that is invariant to %multiplicative 
rescalings of the outcome, and the right-hand side is a product between a sample correlation computed in the pilot sample and a standardized difference defined as follows:
\begin{align}
\label{eqn:ovb_smd}
\text{cross-sample SMD} = \frac{\Bar{X}_{j1, analysis}-\Bar{X}_{j0, analysis}}{S_{X_{j, pilot}}}
\end{align}
The SMD calculates the difference between treated and control groups from the analysis sample and is standardized by the standard deviation from the pilot sample. Hence, we define this SMD as ``cross-sample SMD".

Since the standardized OVB is a product of two terms, level sets for bias take the form of hyperbolic curves on the jointVIP to demarcate equivalent levels of confounding under the crude one-confounder models. In addition, a measure of bias may be computed for any individual variable using its respective SMD and outcome correlation, and color-coding based on these quantities is used for plotting points. We refer to the marginal bias measure as ``unadjusted bias" to distinguish from the typical multivariate OVB models. 

\subsection{{Bias in a finite population framework} \label{subsec:finite_pop_bias}}
{The bias analysis of Section \ref{subsec:bias_curves} assumes  covariates, treatments, and outcomes are sampled jointly from an infinite population.  Although the case study in Section \ref{application} instead uses a finite population framework,
this analysis
still turns out be relevant. 
Given $K$ matched pairs (with the treated unit indexed $k1$ in each pair $k$  and the control unit indexed $k2$), that unobserved confounding is absent, and that $Y_{ki}(1) - Y_{ki}(0) = \tau$ for all $k,i$. The bias of a matched difference-in-means estimator for $\tau$,  viewing only $Z$ as a random variable and holding potential outcomes $Y(1),Y(0)$ and covariates $X$ fixed, can be written as follows:

\begin{align}
\frac{1}{K}\sum^K_{k=1}[Y_{k1}(0) - Y_{k2}(0)](p_{k1} - p_{k2}) \label{eqn:pair_bias}
\end{align}
where $p_{ki} = \frac{\lambda_{ki}/(1-\lambda_{ki})}{\lambda_{k1}/(1-\lambda_{k1}) + \lambda_{k2}/(1-\lambda_{k2})}$ with $\lambda_{ki}$ representing the propensity score for unit $ki$; for derivations see  \citet[\S 4]{sales2018rebar} and 
\citet{huang2022variance}.  This formula suggests that attention to covariate-outcome relationships can improve  estimation and inference via reduction in the magnitude of the $Y_{k1}(1) - Y_{k2}(0)$ terms.  In principle it would vanish if matching were exact on the propensity score, but in practice this is implausible \citep{guo2023statistical, pimentel2023covariate}.   Additionally, if we consider the expected behavior of this bias when potential outcomes are drawn from a model  and covariate $X$ is ignored, we arrive at an approximate bound that is a rescaled version of unadjusted bias (see the Supplemental Appendix A.2 for full derivation).  {Under similar assumptions, \citet{rosenbaum2005heterogeneity} shows that reducing the variance of the $Y_{k1}(0) - Y_{k2}(0)$ terms reduces sensitivity to unmeasured bias, even when propensity score matching is exact.}   
In summary, although in Section 2.2 we did not explicitly motivate the unadjusted bias curves in the context of biases incurred under matched designs nor explicitly invoke the finite-sample framework typically used to analyze such designs, the tools developed in Section 2.2 retain useful interpretations from the perspective of matched analysis. We anticipate similar benefits for other causal inference strategies.}

{\subsection{Using jointVIP to guide study design}
\label{subsec:jointvip_practice}}

{Once the jointVIP has been created, researchers can select variables with large potential bias contributions (as measured by the unadjusted bias curves) for adjustment or otherwise leverage its information to choose tuning parameters.   In a matched study, selected variables might be used to create a Mahalanobis distance \citep{hansen2004full}, or their marginal imbalance could be restricted via fine or refined balance constraints \citep{yang2012optimal, pimentel2015large} as in our case study in Section \ref{application}. In a study using stable balancing weights inverse values of the outcome correlations plotted on the $y$-axis of the jointVIP could be used as balance tolerances \citep{zubizarreta2015stable}. 
In outcome regression settings where the data is too high-dimensional to allow inclusion of all covariates, variables highlighted by jointVIP could be chosen as regressors. 
For matched and weighted studies, a post-design version of the jointVIP can also be created using new SMDs computed on the matched or weighted data. This can suggest further refinements of the original matching or weighting specification, or whether residual bias is large enough to require additional regression adjustment after matching and weighting and which variables should be included in such an adjustment model \citep{rosenbaum2002covariance}.  Table \ref{tab:workflow} summarizes the process of creating and applying jointVIP for practitioners, and a simulation study in Supplemental Appendix A.3 empirically demonstrates the value of this process for bias reduction.}

{A natural question is how or whether to combine the process just described with the balance testing approach to study design proposed by \citet{hansen2008covariate} for matched or stratified observational studies.  Here the design is improved iteratively until an omnibus test using all measured covariates fails to reject the hypothesis that treatment is distributed uniformly within strata. While the jointVIP framework offers important new information by leveraging outcome-covariate relationships ignored by balance tests, balance tests offer a clearer ideal benchmark for success in the form of a hypothetical study randomized within strata, and a single condition to check incorporating all covariates. A researcher might proceed by requiring the final stratified design both to pass a balance test and to minimize potential bias as computed under jointVIP to enjoy the benefits of both frameworks. If this proves impossible, a researcher might instead use jointVIP to select a priority subset of covariates with highest outcome correlation, and search for a design for which the tests of \citet{hansen2008covariate} fail to detect differences with respect to these variables. For an interesting related proposal to use prognostic information to construct a test statistic for balance testing, see \citet{bicalho2022conditional}.}

\begin{table}[!ht]
\begin{tabular}{ll}
\hline
1. choose pilot sample & define pilot sample either as hold-out set from \\
& main analysis sample or from external historical data \\\hline
2. create the jointVIP & fit outcome correlations from the pilot sample \\ 
& and compute SMD from the analysis sample \\ \hline 
3. identify potential confounders & prioritize variables in top right region \\ 
& of the plot and use bias curves to make fine distinctions \\ \hline
4. adjust for confounders & 
create balance constraints (matching or weighting),\\ & 
a covariate distance (matching),  a regressor \\&
 matrix (outcome regression), etc. using chosen variables. 
\\\hline 
5. (optional) plot post-jointVIP & for matching and weighting re-plot with post-design SMD \\ repeat steps 3-5 & repeat as desired \\ \hline \\
\end{tabular}
\caption{{Suggested procedure for use of the joint variable importance plot. As discussed in Section \ref{subsec:jointvip_construction}, the pilot sample typically consists of controls only. See Section \ref{subsec:jointvip_practice} for further details on practical use of jointVIP.
}}
\label{tab:workflow}
\end{table}

{JointVIP can also draw attention to variables with high treatment-model importance but negligible outcome-model importance, sometimes referred to as instrumental variables or prods \citep{pimentel2016constructed}.  Even when all variables could be used for adjustment, it is wise to exclude to such variables since they can inflate unmeasured confounding bias \citep{brooks2013squeezing, ding2017instrumental}. JointVIP  enables either excluding such variables or (if it is not entirely clear whether a variable should be excluded) constructing multiple control groups that adjust for these variables differently \citep{pimentel2016constructed}.}

{Some caution should be exercised when using and interpreting jointVIP. Outcome correlations can change substantially when variables are transformed; outliers may also skew the means of either treatment or control groups and hence the standardized mean differences. Blindly using all variables above a bias cutoff may also be suboptimal. For example, if two variables are near-perfectly collinear, both would be highlighted as priorities in jointVIP, but adjusting for one may be sufficient to remove bias. Finally, baseline variables that are absent or rare in the pilot sample may not be well-represented in the plot.}

\section{CASE STUDY}
\label{application}

\subsection{Glyburide as a treatment for gestational diabetes}

Due to improved ease of use and lower cost, oral antidiabetic medications, such as glyburide, are often prescribed compared to the recommended insulin therapy as treatment for gestational diabetes \citep{castillo2014trends}. The safety of glyburide, however, remains contentious due to potential transfer to the fetus through the placenta \citep{bulletins2018acog}. The question remains: does glyburide increase the risk of adverse perinatal outcomes in real-world settings? We investigate glyburide's impact on C-section delivery compared to medical nutritional therapy, the universal first-line therapy in a large, population-based cohort. 

The study population consists of Kaiser Permanente Northern California (KPNC) members. Individuals who are diagnosed with GDM receive medical nutritional therapy (MNT) as the universal first line of therapy. Pharmacologic treatment, including oral antidiabetic medications (glyburide, metformin, or other) and/or insulin, is prescribed in addition to MNT if glycemic control goals are not met. Individuals with GDM who received MNT alone constituted our control group while those who additional received glyburide as the only pharmacologic therapy constituted our treatment group. There are 54 common variables between the 2007-2010 data (pilot sample) and 2011-2021 data (analysis sample), including indicators of missing data as variables. Table \ref{tab1} summarizes selected baseline variables (see Supplemental Appendix A.4 for the full data summary). Missing values were imputed separately for each year using random forest \citep{stekhoven2012missforest}. Details about the pattern of missing values and the imputation procedure are reported in Supplemental Appendix A.5. Our use of KPNC data for this study is approved by the KPNC Institutional Review Board, which waived the requirement for informed consent from participants.

\begin{table}
\caption{Summary of selected baseline variables for pregnant individuals with gestational diabetes. \newline {\footnotesize BMI: body mass index, C-C: Carpenter-Coustan, GDM: gestational diabetes, OGTT: oral glucose tolerance test, SD: standard deviation A normal OGTT fasting blood glucose level is lower than 95 mg/dL. Abnormal indicates a result higher than normal.}}
\centering
\begin{adjustbox}{width=1\textwidth}
\setstretch{1}
\begin{tabular}{lllll}
   \hline
   &  & 2007-2010 & 2011-2021 & 2011-2021 \\
   &  & Control & Control & Treated \\
    &  & n = 7,526 & n = 19,183 & n = 10,786 \\
   \hline
   \vtop{\hbox{\strut Family history of}\hbox{\strut \hspace{0.2cm} diabetes = yes (\%)}} &  & 198 (2.6) & 1,202 (6.3) & 822 (7.6) \\
\vtop{\hbox{\strut OGTT for}\hbox{\strut \hspace{0.2cm} fasting blood glucose = abnormal (\%)}} &  & 2,165 (28.8) & 3,762 (19.6) & 4,512 (41.8) \\
  GDM severity = severe (\%) &  & 775 (10.3) & 1,744 (9.1) & 1,215 (11.3) \\
  \vtop{\hbox{\strut GDM diagnosed by the}\hbox{\strut \hspace{0.2cm} C-C criteria = yes (\%)}} &  & 7,323 (97.3) & 17,303 (90.2) & 8,419 (78.1) \\
  Age (\%) & Under 25 & 552 (7.3) & 1,029 (5.4) & 349 (3.2) \\
   & Between 25-29 & 1,665 (22.1) & 3,762 (19.6) & 1,866 (17.3) \\
   & Between 30-34 & 2,584 (34.3) & 7,101 (37.0) & 4,165 (38.6) \\
   & Over 35 & 2,725 (36.2) & 7,291 (38.0) & 4,406 (40.8) \\
  \vtop{\hbox{\strut Gestational age at}\hbox{\strut \hspace{0.2cm} GDM diagnosis (mean (SD))}} &  & 26.06 (6.10) & 26.86 (6.02) & 23.53 (7.12) \\
  History of macrosomia = yes (\%) &  & 69 (0.9) & 145 (0.8) & 147 (1.4) \\
  History of GDM = yes (\%) &  & 856 (11.4) & 3,401 (17.7) & 2,608 (24.2) \\
  Parity (\%) & 0 & 3,121 (41.5) & 7,611 (39.7) & 3,873 (35.9) \\
   & 1 & 2,347 (31.2) & 6,566 (34.2) & 3,994 (37.0) \\
   & more than 2 & 2,058 (27.3) & 5,006 (26.1) & 2,919 (27.1) \\
  Pre-pregnancy BMI (\%) & Underweight & 100 (1.3) & 391 (2.0) & 76 (0.7) \\
   & Normal & 1,921 (25.5) & 5,107 (26.6) & 1,706 (15.8) \\
   & Overweight & 2,847 (37.8) & 6,369 (33.2) & 3,361 (31.2) \\
   & Obese & 2,658 (35.3) & 7,316 (38.1) & 5,643 (52.3) \\
  Race/ethnicity (\%) & \vtop{\hbox{\strut Asian or}\hbox{\strut \hspace{0.1cm} Pacific Islander}}& 2,919 (38.8) & 8,553 (44.6) & 4,560 (42.3) \\
   & Hispanic & 2,322 (30.9) & 5,013 (26.1) & 2,923 (27.1) \\
   & White & 1,602 (21.3) & 4,090 (21.3) & 2,381 (22.1) \\
   & \vtop{\hbox{\strut Black or}\hbox{\strut \hspace{0.1cm} African American}} & 315 (4.2) & 736 (3.8) & 410 (3.8) \\
   & Other or unknown & 368 (4.9) & 791 (4.1) & 512 (4.7) \\
  \vtop{\hbox{\strut Pre-pregnancy}\hbox{\strut \hspace{0.2cm} pre-diabetes = yes (\%)}} &  & 479 (6.4) & 1,802 (9.4) & 1,915 (17.8) \\
  \vtop{\hbox{\strut Glucose challenge}\hbox{\strut \hspace{0.2cm} test value (mean (SD))}} &  & 169.43 (22.38) & 169.71 (22.14) & 173.22 (24.32) \\
   \hline \\
\end{tabular}
\end{adjustbox}
\label{tab1}
\end{table}

\subsection{Design}

\subsubsection{Variable selection using jointVIP}

JointVIP is constructed {using the} \texttt{jointVIP} package  in R; for a brief software tutorial see \citet{liao2023jointvip}. To ensure particularly stringent control of the propensity score, we impose a caliper equal to 0.2 standard deviations of the fitted propensity score values in the entire sample. Using a caliper on the propensity score is  a natural choice because our approach to inference relies on similar propensity scores within matched pairs \citep{pimentel2023covariate}. We match exactly on year to address substantive concerns about potential for temporal shifts in the standard of care in the absence of reliable outcome correlations. 

We address potential bias from additional variables by imposing a series of refined balance constraints tailored to the outcome. Refined covariate balance enables users to specify top-priority variables and their interactions to be balanced as though they were the only variables in the study, with lower-priority variables receiving further attention as possible \citep{pimentel2015large}. While this framework offers substantial flexibility to the researcher, it relies on strong substantive knowledge to specify the balance tiers in a reasonable manner. Frequently it is not immediately clear how to organize a group of baseline variables into balance tiers in a principled way. JointVIP offers a data-driven approach in settings where ambiguity remains even after accounting for substantive knowledge. We specify tiers of variables for refined covariate balance by identifying sets of variables with high importance. Since the prognostic score (fit in the pilot sample using LASSO regression) ranks among the variables contributing the largest unadjusted bias, we include quintiles of the prognostic score in the first balance tier. We include all variables contributing unadjusted bias greater than or equal to 0.010 with variables in subsequent tiers, with those contributing larger amounts of bias in higher tiers. Table \ref{tab:tiers} summarizes the chosen balance tiers for the design. Specific potential bias values can be found in Supplemental Appendix A.6 column \textit{Pre-matched bias}. In addition, we discretized the continuous variable for gestational age at GDM diagnosis for compatibility with refined covariate balance algorithm. 

\begin{table}[!ht]
\center
\begin{tabular}{cl}
Balance tier & C-section delivery \\\hline
1 & Prognostic score quintile \\\hline
2 & OGTT for fasting blood glucose  \\
  & Obese pre-pregnancy BMI \\\hline
3 & GDM diagnosed by Carpenter-Coustan criteria \\
& Gestational age category at GDM diagnosis \\
& Pre-pregnancy pre-diabetes \\
& Normal pre-pregnancy BMI \\\hline \\
\end{tabular}
\caption{Balance tiers for refined covariate balance for each outcome, chosen using jointVIP plots. \newline {\footnotesize BMI: body mass index, C-section: Cesarean section, GDM: gestational diabetes,
jointVIP: joint treatment-outcome variable importance plot, OGTT: oral glucose tolerance test.}}
\label{tab:tiers}
\end{table}

\begin{figure}[!hb]
    \begin{center}
        \includegraphics[scale=0.65]{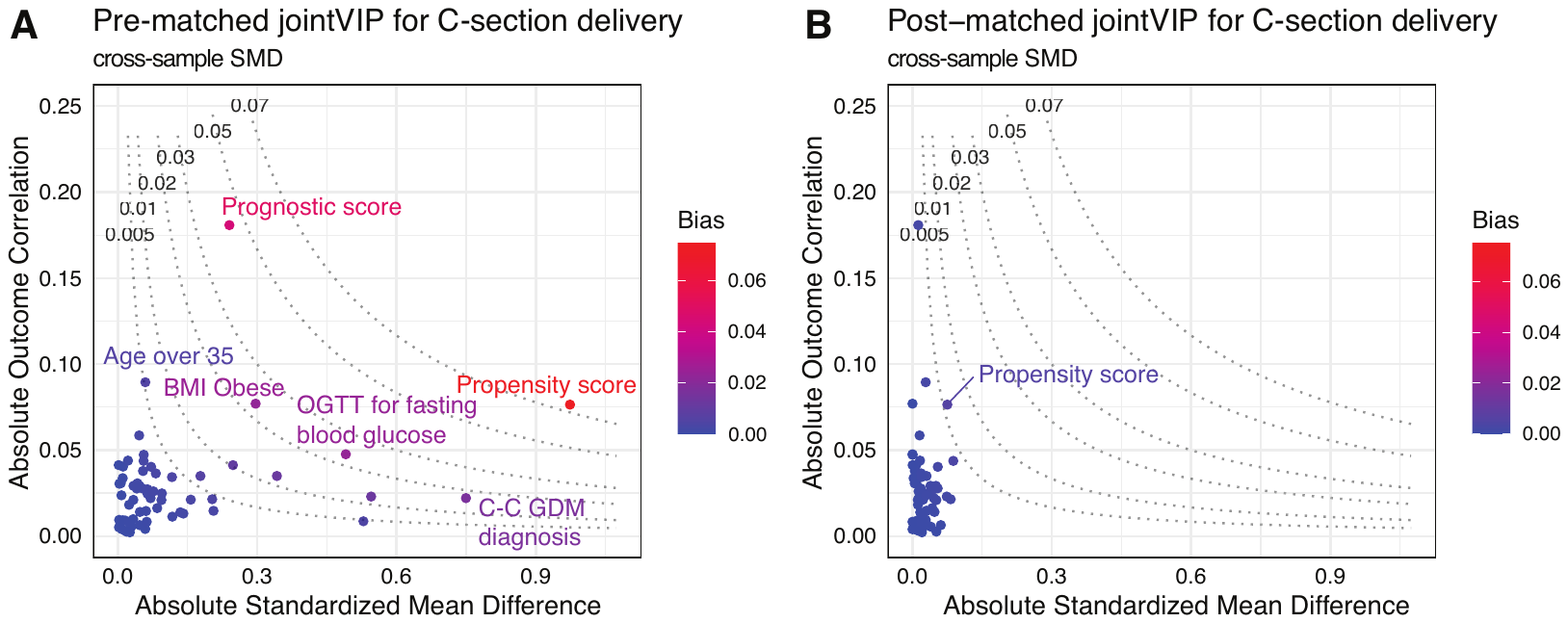}
    \end{center}
    \caption{Pre-post match results for Cesarean section delivery. \newline {\footnotesize BMI: body mass index, C-section: Cesarean section, GDM: gestational diabetes, jointVIP: joint treatment-outcome variable importance plot, OGTT: oral glucose tolerance test, SMD: standardized mean difference. }}
    \label{fig:cdflag}
\end{figure}

\subsubsection{Matched Design}

We conduct matching with refined covariate balance using the \texttt{rcbalance} package in R and the balance tiers in Table \ref{tab:tiers}. Post-matched jointVIP results, reflecting new levels of balance after matching, are plotted in Figure \ref{fig:cdflag}.B. For variables that were specified, post-matched biases are compared to pre-matched biases in Supplemental Appendix A.6, which shows all baseline variables and summary measures to have small biases (around 0.005 or less) post-matching. Note in particular that variables with high outcome correlation are balanced especially well, a feature of the design that traditional methods based on Love plots are not equipped to guarantee.

2,093 treated subjects are excluded from the match due to caliper and exact matching constraints, and 8,693 pairs are matched. Those who are excluded tend to have more signs of severe GDM and higher probability of treatment; it is not surprising that it is difficult to find comparable controls for matching them (Supplemental Appendix A.7). We note that the average risk difference for C-section is best understood not as an estimate of an average treatment effect on the treated \citep{stuart2010matching}, but as an average effect on a ``marginal" population consisting of individuals for whom treatment by either arm is reasonably likely \citep{rosenbaum2012optimal, %li2013weighting, 
li2019addressing, greifer2021choosing}. This estimand, while less common in theoretical discussions of causal inference, adheres more closely to the substantive quantity of interest for physicians who are typically more interested in guidance for patients with equipoise, and less interested in effects on patients who would clearly be assigned glyburide or not in the large majority of cases.

\subsection{{Outcome analysis}}
\label{inference}
%\subsubsection{{Randomization inference framework}}
To perform inference, we index matched pairs by $i = 1, \cdots, I$, and individuals in each matched pair by $k = 1, 2$. For a matched pair $i$, one person is treated with glyburide, $Z_{ik} = 1$, and the other with MNT, $Z_{ik} = 0$, hence $Z_{i1} + Z_{i2} = 1$. Let $\mathcal{Z}$ denote the event that $Z_{i1} + Z_{i2} = 1$ for each matched pair $i$. Each subject $ik$ has corresponding potential outcomes $Y_{{ik}}(1)$ and $Y_{{ik}}(0)$ for treatment with and without glyburide respectively. We collect quantities fixed in advance of treatment, including potential outcomes and covariates, in the set $\mathcal{F} = \{(Y_{{ik}}(1), Y_{{ik}}(0), \boldsymbol{x}_{ik}), i=1,\cdots, I, k = 1,2\}$. Our outcome of interest is a binary indicator for C-section.

We test the %\citet{fisher1936design} 
sharp null hypothesis, $H_0: Y_{{ik}}(1) = Y_{{ik}}(0)$ for all $i,k$. Assuming that paired subjects are equally likely to receive glyburide, we can test this hypothesis by repeatedly permuting treatment indicators within pairs (independently across pairs) with probability ${1}/{2}$; this corresponds to resampling treatment indicators conditional on $\mathcal{Z}$ and $\mathcal{F}$. Since under the sharp null the outcomes remain identical regardless of treatment assignment, we can compute a test statistic under each permutation using observed outcomes and compare the actual observed value of the test statistic to this reference distribution to conduct inference. For binary outcomes, in particular, we may apply McNemar's test \citep{mcnemar1947note}. 
The above procedure relies on the assumption $Pr(Z_{ik}=1|\mathcal{F}, \mathcal{Z}) = {1}/{2}$ with independent assignment for each pair, which is true when unobserved confounding is absent and propensity scores are matched exactly; {it is a quasi-randomization test in the sense of \citet{zhang2022randomization}}. In real observational studies this assumption may fail, and sensitivity analysis is needed to probe the robustness of the initial findings to such failures. We perform sensitivity analysis as described in \citet{rosenbaum2010design} Section 3. 

\subsubsection{Results}

There are $2 \times 8,693$ individuals who are matched in pairs, 6,023 (34.64\%) individuals delivered by C-section. Matched results are shown in Table \ref{tab_matched_csec}. For control (MNT only) individuals, 33.61\% delivered by C-section, and for treated (glyburide and MNT) individuals, 35.67\% delivered by C-section (raw treatment-control difference of 2.06\%). McNemar's test yields a one-sided p-value of 0.0020. Evaluating at significance level 0.05, there is evidence to reject the null hypothesis under a no unmeasured confounding assumption. However, the sensitivity analysis produces a threshold $\Gamma$ of 1.041, which indicates that a very small degree of unmeasured confounding (the amount needed to shift a a 0.50 probability of treatment to a  1.041/(1 + 1.041) $\approx$ 0.51 probability of treatment) can explain away the causal effect detected. As such we find no substantial evidence that glyburide is causing the increase in cases of C-section delivery in this study.

\begin{table}
\center
\begin{tabular}{clcc}
\multicolumn{4}{r}{\textbf{Treated with glyburide}} \\\cmidrule{3-4}
& & {C-section} &  {not C-section}  \\\cmidrule{3-4}
{\textbf{Control}} & {C-section}  &   1078 & 1844  \\\cmidrule{2-4}
& {not C-section}            &   2023 & 3748  \\\cmidrule{2-4} \\
\end{tabular}
\caption{Matched analysis for Cesarean section delivery.}
\label{tab_matched_csec}
\end{table}

\section{DISCUSSION}
\label{discussion}

JointVIP is a useful tool for selecting variables to balance during the observational study design phase. One notable advantage over traditional methods is the visual ease of comparison for marginal relationships of each variable with both the outcome and treatment. 
Methods leveraging jointVIP {can} offer better bias reduction and increased robustness against unmeasured confounders \citep{rosenbaum2005heterogeneity}. 
Several other authors have discussed ideas closely related to jointVIP. 
\citet{zhao2022outcome} propose variable selection for fitting generalized propensity scores using measures of outcome importance and provide supporting theory suggesting the optimality of this approach. \citet{aikens2020pilot} and \citet{aikens2022assignment} construct an alternative design-stage visualization based partially on a pilot sample incorporating outcomes, the assignment-control (AC) plot. In contrast to jointVIP however, the AC plot represents subjects rather than variables on the plot, using the estimated prognostic score and propensity score values on the axes. AC plots and jointVIP thus provide valuable complementary representations of observational data. 
{ Finally, \citet{cinelli2020making} propose a similar contour plot based on omitted-variable-bias calculations that consider each variable in turn as a potential omitted confounder, for use in interpreting parameters in sensitivity analysis.  
For matching and weighting, the post-match jointVIP has potential to be used in a similar way.  However, additional mathematical work is required to establish a mapping between the $\Delta_j$ and $\beta_j$ quantities represented on the jointVIP and the parameters of existing sensitivity analysis approaches.}

A natural question is why the omitted variable biases for the unadjusted bias curves should be computed under the one-covariate model in equation (\ref{eq:eq1}) instead of a model containing all measured covariates. This relates to a larger question about whether to focus on visualizing marginal measures of association between covariates and treatment or outcome, or instead to focus on conditional or partial measures that account for other variables. We focus on marginal measures 
rather than conditional measures (such as multiple regression coefficients from models for treatment or outcome and OVB from excluding one variable from a regression with many covariates), in contrast to previous works such as \citet{cinelli2020making}. While previous authors focused on post-hoc sensitivity analyses in which a single model had already been chosen for analysis, jointVIP is a {pre-analysis} tool aimed at helping select covariates for which to adjust. As such, it is unclear which covariates should be adjusted for in computing partial correlations with outcome and treatment. This is especially true in high-dimensional settings where the number of covariates may exceed the number of sample points in either the pilot or main analysis sample, in which case partial measures of association may not be well-defined for some sets of adjustment covariates. We also note that current standard heuristics 
emphasize reporting and minimizing SMDs rather than regression coefficients from a propensity score, so a marginal approach generalizes existing practice more naturally (as demonstrated above).
{
However, developing a conditional jointVIP is an interesting topic for future work.  For example, a forward-selection method with attention to multicollinearity could be developed by selecting only one variable for adjustment from the original jointVIP, then creating a conditional version of jointVIP for the remaining variables where all plotted measures adjust for the first selected variable, and iterating until a stopping criterion is reached.}

{While we focused on using pilot samples consisting only of controls, if extensive treatment effect heterogeneity is present this approach might underestimate the bias contributed by individual variables.  Instead, one could take a pilot sample from each study arm and fit distinct treatment and control outcome correlations $\beta_j^{(1)}$ and $\beta_j^{(0)}$.  A generalized version of our argument in Section \ref{subsec:bias_curves} due to \citet{zhao2021covariate} suggests plotting $\beta_j^{(1)}p_0 +\beta_j^{(0)}p_1$ on the y-axis of the jointVIP, where $p_1$ and $p_0$ are the anticipated proportions of treated and control subjects in the final design.  Of course, it may not be advisable to sacrifice treated subjects to the pilot sample for such an analysis when treatment is rare.
}

{Another area for future work is generalizing jointVIP to allow for nonlinearity. Pearson correlation captures linear relationships but may miss strong nonlinear relationships.  Nonlinear measures of importance 
such as the interpretable 
mean decrease in impurity (MDI+)
derived by \citet{agarwal2023mdi} for random forests, could in principle be used on the y-axis of the jointVIP. 
Two primary challenges arise.  First is the question of marginal versus conditional relationships raised above, if nonlinear importance measures vary depending on the other variables included in the model. 
 Second is the difficulty of deriving nonlinear versions of the unadjusted bias curves.  
 Statistical interpretation of variable importance in nonlinear models such as random forest is an active research area and we are not aware of any straightforward generalization of omitted variable bias for this context.}

\newpage
\section*{Acknowledgments}
The authors gratefully acknowledge support from \textit{Hellman Fellowship, National Science Foundation 2142146 and DGE 2146752, National Institute of Diabetes and Digestive and Kidney Diseases K01DK120807, National Heart, Lung, and Blood Institute R01HL157666, and Kaiser Permanente Northern California Community Benefits Program RNG209492}. The authors thank {David Bruns-Smith,} Avi Feller, {Erin Hartman,} Melody Y. Huang, {Yaxuan Huang, Sizhu Lu, Arisa Sadeghpour, Andy Shen, and Arnout van Delden for valuable comments.

\section*{Conflict of interest}
The authors report there are no competing interests to declare.

\bibliography{biblio}

\end{document}